# Transition from Sign-reversed to Sign-preserved Cooper-pairing Symmetry in Sulfur-doped Iron Selenide Superconductors

Qisi Wang[1], J. T. Park[2*], Yu Feng[1], Yao Shen[1], Yiqing Hao[1], Bingying Pan[1], J. W. Lynn[3], A. Ivanov[4], Songxue Chi[5], M. Matsuda[5], Huibo Cao[5], R. J. Birgeneau[6,7], D. V. Efremov[8], Jun Zhao[1,9†]

[1]*State Key Laboratory of Surface Physics and Department of Physics, Fudan University, Shanghai 200433, China*

[2]*Heinz Maier-Leibnitz Zentrum (MLZ), Technische Universitat Munchen, D-85748 Garching, Germany*

[3]*NIST Center for Neutron Research, National Institute of Standards and Technology, Gaithersburg, Maryland, 20899, USA*

[4]*Institut Laue-Langevin, 71 Avenue des Martyrs, 38042 Grenoble Cedex 9, France*

[5]*Quantum Condensed Matter Division, Oak Ridge National Laboratory, Oak Ridge, Tennessee 37831-6393, USA*

[6]*Department of Physics, University of California, Berkeley, California 94720, USA*

[7]*Department of Materials Science and Engineering, University of California, Berkeley, California 94720, USA*

[8]*IFW Dresden, Helmholtzstr. 20, 01069 Dresden, Germany*

[9]*Collaborative Innovation Center of Advanced Microstructures, Fudan University, Shanghai 200433, China*

**Abstract**

An essential step toward elucidating the mechanism of superconductivity is to determine the sign/phase of superconducting order parameter, as it is closely related to the pairing interaction. In conventional superconductors, the electron-phonon interaction induces attraction between electrons near the Fermi energy and results in a sign-preserved *s*-wave pairing. For high-temperature superconductors, including cuprates and iron-based superconductors, prevalent weak coupling theories suggest that the electron pairing is mediated by spin fluctuations which lead to repulsive interactions, and therefore that a sign-reversed pairing with an $s_{\pm}$ or *d*-wave symmetry is favored. Here, by using magnetic neutron scattering, a phase sensitive probe of superconducting gap, we report the observation of a transition from the sign-reversed to sign-preserved Cooper-pairing symmetry with insignificant changes in $T_c$ in the S-doped iron selenide superconductors $K_xFe_{2-y}(Se_{1-z}S_z)_2$. We show that a rather sharp magnetic resonant mode well below the superconducting gap ($2\Delta$) in the undoped sample ($z = 0$) is replaced by a broad hump structure above $2\Delta$ under 50% S doping. These results cannot be readily explained by simple spin fluctuation-exchange pairing theories and, therefore, multiple pairing channels are required to describe superconductivity in this system. Our findings may also yield a simple explanation for the sometimes contradictory data on the sign of the superconducting order parameter in iron-based materials.

PACS numbers: 74.25.Ha, 74.25.Dw, 74.70.-b, 78.70.Nx

The prevailing paradigm of iron-based superconductors consists of the concept of Cooper pairing to a great degree due to interband interactions, stemming from spin fluctuations [1]. This belief is supported by several studies. For instance, iron-based superconductivity exists in the vicinity of an antiferromagnetic (AFM) ordered ground state [2]; strong spin fluctuations whose momentum structure matches the Fermi surface geometry are obtained theoretically and documented experimentally [1-4]. Moreover, *ab initio* calculations show that the electron-phonon coupling is very small [5]. The main consequence of the domination of the interband coupling due to spin fluctuations is the existence of sign-reversed pairing symmetry. Magnetic neutron scattering is an important tool to probe the relative phase of the superconducting order parameter [2,6,7]. It was predicted theoretically and later observed experimentally, that a sharp resonance peak in the dynamical spin susceptibility $\chi''(Q,\omega)$ (magnetic resonance mode) may appear in the superconducting state at energy below 2Δ, due to creation particle-hole pairs. This process is characterized by the coherence factor $\left(1 - \frac{\epsilon_q \epsilon_{q+Q} + \Delta_q \Delta_{q+Q}}{E_q E_{q+Q}}\right)$ [8]. Here $\Delta_q$ is the gap function, $\epsilon_q$ the single particle dispersion and $E_q$ the quasiparticle energy. The coherence factor vanishes if the gap function has the same signs at the momenta at which the particle-hole pairs are created $\Delta_q = \Delta_{q+Q}$, while it has a finite values for $\Delta_q = -\Delta_{q+Q}$. Hence emergence of the resonance mode in the superconducting state is a strong evidence of a sign reversed order parameter [1-3,6,7,9-12]. Recently it was pointed out that in the case of the sign preserved order parameter, there is a redistribution of the magnetic spectral weight below $T_c$ due to the opening of superconducting energy gap in the spin excitation channel, which leads to a nonresonance broad peak above 2Δ [13-15]. Therefore the observation of the peak either below or above 2Δ allows identifying the relative sign of the order parameter at the propagation vector. For example, the electron-pairing symmetry in cuprates is ubiquitously *d* wave, which is supported by the existence of a sharp magnetic resonant mode at the "hot spot" wave vector below 2Δ [1,6,16]. In the case of iron-based superconductors, although no such universal pairing symmetry has been identified, similar resonant modes have also been observed in many systems [2]. For most iron pnictide superconductors, the resonant mode appears at the nesting wave vector (π, 0) between hole Fermi surfaces at the zone center and electron Fermi surfaces at the zone edge (1-Fe unit cell), indicating an *s*-wave pairing with sign reversal gap function ($s_\pm$) between electron and hole Fermi surfaces [3,17]. For the recently discovered alkali-metal-intercalated iron-selenide superconductors $A_x$Fe$_{2-y}$Se$_2$ (A = K, Rb ...) with no hole Fermi surfaces at the zone center, the resonance wave vector matches the nesting wave vector (π, 0.5π) between electron Fermi surfaces at two adjacent zone edges, suggesting a *d*-wave or other types of $s_\pm$-wave pairing [18-28]. Although the presence of a resonant mode is broadly consistent with a spin-fluctuation mediated sign-reversed pairing symmetry, several issues remain to be settled. First, the resonant modes in some iron-based superconductors are broader than predicted in theories and appear at energy close to the superconducting gap edge where the requirements of the resonant mode as a bound state below 2Δ cannot be unambiguously confirmed [2]. Second, a few measurements yield contradicting results regarding the phase of the superconducting order parameter in different materials [12,18,29,30].

We use neutron scattering to study the effect of isovalent S doping on the spin dynamics and its relationship with superconductivity in K$_x$Fe$_{2-y}$(Se$_{1-z}$S$_z$)$_2$. Single crystals of four compositions of K$_x$Fe$_{2-y}$(Se$_{1-z}$S$_z$)$_2$ ($z = 0$, $T_c = 31.2$ K; $z = 0.25$, $T_c = 32.0$ K; $z = 0.4$, $T_c = 28.4$ K; $z = 0.5$, $T_c = 25.4$ K) were prepared by the self-flux method. The superconducting properties of four single crystals were characterized with DC magnetic susceptibility and resistivity measurements on small pieces of crystals cut from large single crystals used for neutron scattering experiments [31]. Our neutron diffraction refinements suggest that the single crystals are phase separated into a √5×√5 iron vacancy ordered block AFM insulating phase and a superconducting phase with a 122 type tetragonal crystal structure without static magnetic order, consistent with earlier measurements [32]. S-doping monotonically reduces the lattice constants and Fe-Se/S bond distances but does not change the symmetry of the crystal structure. Actual chemical compositions are determined by neutron and X-ray diffraction refinements. We note that the K and Fe concentrations do not evolve rapidly under S doping [31], which is consistent with the ARPES measurements on the same batch of crystals that the carrier concentration displays little change for all S doped samples measured [33]. Inelastic neutron scattering measurements were carried out on the HB3 and HB1 thermal triple-axis spectrometers at the High-Flux-Isotope Reactor (HFIR), Oak Ridge National Laboratory, United States, the IN8 thermal triple-axis spectrometer at the Institute Laue-Langevin, Grenoble, France, the PUMA thermal triple-axis spectrometer at the Heinz Maier-Leibnitz Zentrum (MLZ), Technische Universität München, Garching, Germany, and the BT-7 thermal triple-axis spectrometer at the NIST Center for Neutron Research (NCNR), United States. The single crystals were aligned in the ($H$, $K$, 0) scattering plane within ~ 1.5 degrees mosaicity for the measurements. We define the wave vector **Q** at ($q_x$, $q_y$, $q_z$) as ($h$, $k$, $l$) = ($q_x a/2\pi$, $q_y b/2\pi$, $q_z c/2\pi$) reciprocal lattice units (r.l.u.) in the 1-Fe unit cell.

Figure 1 shows the detailed temperature dependent energy scans at **Q** = (-0.5, 0.77, 0) for the undoped sample ($z = 0$, $T_c = 31.8$ K). A prominent enhancement of the scattering at around 13 meV is observed in the spin excitation spectrum of the superconducting state [Fig. 1(a)], which is similar to the resonant modes reported previously [18,19]. This can be illustrated more clearly by eliminating the energy-dependent background, by subtraction of the signal at $T = 36$ K from that in the superconducting state. As shown in Fig. 1(b), the background subtracted contour map exhibits a sharp resonant excitation (13 meV), well below the superconducting gap ($2\Delta = 20.6$ meV) at $T = 4$ K. More interestingly, both the resonant mode ($E_p$) and superconducting gap $2\Delta$ hardly soften on heating and undergo a sharp transition at $T_c$. This behavior differs from the temperature dependence of a conventional weak coupling BCS superconducting gap, which further implies the unconventional nature of superconductivity in $K_x Fe_{2-y} Se_2$.

To determine the in-plane momentum structure of the spin excitations as a function of S-doping, we performed **Q**-scans for the $z = 0$, 0.25, 0.4, and 0.5 samples below and above the superconducting transition temperature $T_c$. In the $z = 0$ sample, strong magnetic excitations are observed in the normal state near **Q** = (-0.5, 0.77, 0) [Fig. 2(a)], which is close to the commensurate wave vector **Q** = ($\pi$, 0.5$\pi$, 0) reported in $A_x Fe_{2-y} Se_2$ [18,19,28]. Interestingly, the magnetic wave vector barely changes with increasing S concentration up to 50% [Figs. 2(b)-2(d)]. Such doping independent magnetic wave vector is in line with the ARPES measurements that Fermi surface geometry in $K_x Fe_{2-y}(Se_{1-z}S_z)_2$ remains essentially unchanged up to 80% S doping [33]. For all samples measured, the magnetic excitations are enhanced on cooling below $T_c$. Although the $T_c$ is suppressed very slowly upon S doping, the increase of the magnetic signal in the superconducting state is clearly less pronounced at higher S concentrations.

Figure 3 illustrates the evolution of the resonant mode as a function of S concentration. The background subtracted energy scan (5 K-36 K) of the undoped sample ($z = 0$) shows a sharp resonant mode, below which a spin gap is also observed due to the opening of the superconducting gap [Fig. 3(a)]. We note that the magnitude of the spin gap ($\Delta_s \approx 10.2$ meV) is much smaller than that of the superconducting gap because of the presence of the in-gap resonant mode. The most interesting observation is that in the $z = 0.25$ sample, in addition to the resonant excitation below $2\Delta$, the spectra exhibit a weak shoulder on the higher energy side above $2\Delta$ [Fig. 3(b)]. With increasing S concentration to $z = 0.4$, the relative intensity of the resonant mode decreases as the shoulder becomes more prominent [Fig. 3(c)], and eventually the resonant mode is completely replaced by a broad hump structure above $2\Delta$ in the $z = 0.5$ sample [Fig. 3(d)]. The magnitude of the spin gap at $z = 0.5$ is very close to $2\Delta$, indicating that the scattering below the superconducting gap is eliminated. The absence of magnetic resonant mode below $2\Delta$ and the pile up of states above $2\Delta$ suggest that the superconducting order parameter no longer has the opposite sign between two adjacent electron Fermi surfaces at this doping level. Indeed, a similar magnetic fluctuation spectrum has been predicted in iron pnictides assuming that the superconducting gap in electron and hole Fermi surfaces possesses the same sign [13,15].

To confirm that the enhancement of the spin excitations in the superconducting state is truly associated with superconductivity, we have measured the detailed temperature dependence of the resonant mode and the hump structure above 2Δ [Figs. 3(e)-3(h)]. The scattering of the resonant mode at 13 meV displays an order-parameter-like behavior with an onset at $T_c$, which proves a tight connection between them [1,2,6]. Moreover, a similar temperature dependence of the scattering of the hump/shoulder structure above 2Δ is also observed [Figs. 3(j), 3(k) and 3(h)], suggesting that the redistribution of the spectral weight across 2Δ is coupled with the opening of the superconducting gap.

The doping dependence of the $T_c$, superconducting gap (2Δ), and superconductivity-induced magnetic peak energy ($E_p$) are summarized in Fig. 4. It is shown that the $T_c$ exhibits a small, but observable, increase from 31.2 K at $z = 0$ to 32.0 K at $z = 0.25$ followed by a gradual decrease to 25.4 K at $z = 0.5$ [Fig. 4(a)]. This is consistent with the phase diagram reported previously [34]. On the other hand, the 2Δ determined by ARPES measurements on the same batch of samples shows a monotonic gradual decrease with increasing S concentration [33]. The magnetic peak energy ($E_p$) barely changes from $z = 0$ to $z = 0.4$ and suddenly shifts to higher energy at $z = 0.5$. For $z = 0, 0.25, 0.4$, the ratio $E_p/2\Delta$ is close to the empirical universal ratio of ~0.64 for magnetic superconductors [35], which meets the requirement that the resonant mode is a bound state below the superconducting gap in a sign-reversed pairing state. This ratio jumps to 2.03 at $z = 0.5$, which obviously violates the primary criteria of the magnetic resonant mode [Fig. 4(b)]. It is interesting to note that S-doping changes the pairing symmetry, while it has an insignificant impact on $T_c$. Two components in spin excitation spectra, i.e., the sharp resonant mode below the 2Δ and hump structure above 2Δ, are clearly observed at $z = 0.25$ and 0.4. It is worth to note that ARPES and transport measurements have shown that S doping increases the bandwidth and reduces the electron correlation considerably but does not change the carrier concentration or Fermi surface geometry, and that the superconducting gap remains essentially isotropic under S-doping [33,34,36].

We now discuss the possible mechanisms by which S-doping may influence the symmetry of the gap function. First we discuss the proposal that S-doping acts as nonmagnetic impurities. In contrast to cuprates where the superconductivity is rapidly suppressed by a small amount of nonmagnetic impurities in the CuO2 plane [16], iron pnictides are much more robust against nonmagnetic impurities [37]. In $K_xFe_{2-y}Se_2$, superconductivity survives up to 70% isovalent S-doping [33,34]. This seems to agree with theoretical calculations, which show that the robustness of $T_c$ is due to the multiband electron structure and the effects of strong scattering [38-41]. Moreover, nonmagnetic impurities may cause a transition from the $s_\pm$ pairing to $s_{++}$ pairing [38,39]. If this applies in $K_xFe_{2-y}(Se_{1-z}S_z)_2$, one would expect stabilizing of the critical temperature after the transition, which was not observed by the transport and ARPES measurements [33,34]. Moreover, ARPES measurements show that spectral weight distribution in momentum space has not been disturbed by S-doping significantly, indicating that the effect of impurity scattering is negligible [33]. These results suggest that the impurity effect is unlikely the predominant underlying reason for the phase transition.

On the other hand, it has been suggested that a conventional sign-preserved s-wave pairing could be realized in the case of dominant orbital fluctuations which compete with spin fluctuations [14,26]. Therefore one may naively suppose that for the low S-doping spin fluctuations are dominant, while for higher S-doping orbital fluctuations become more important. But in this situation a decrease of $T_c$ may be expected in the intermediate region, as the spin-fluctuations are compensated by the orbital fluctuations. Contrary $K_xFe_{2-y}(Se_{1-z}S_z)_2$ shows maximum of $T_c$ at $z = 0.25$. Hence more sophisticated mechanism, which involves strong intraband interaction and orbital degree of freedom should be considered. One of the possibilities is the novel orbital-selective pairing state which has a conventional $s$-wave form factor with a $d$-wave ($B_{1g}$) pairing symmetry [27,42]. In principle, in this scenario, pure conventional $s$-wave or $d$-wave pairing could be achieved by varying the orbital selectivity of the pairing interactions. A decrease of the orbital-dependent renormalization in the $d$ orbitals is indeed observed in S-doped $Rb_xFe_{2-y}Se_2$ [36], although it is currently unclear how S-doping could change the orbital selectivity of the exchange couplings in this system.

To conclude, we have reported experimental evidence of a transition from the sign-reversed to sign-preserved pairing symmetry in $K_xFe_{2-y}(Se_{1-z}S_z)_2$ iron-based superconductors. Superconductivity is relatively strong in terms of $T_c = 25.4$ K under 50% S doping, where the sign-preserved pairing is dominant. This is difficult to explain by simple weak coupling spin fluctuation-exchange pairing theories. It is possible that $K_xFe_{2-y}(Se_{1-z}S_z)_2$ is close to an intermediate pairing state. The predominant effect of S-doping is to decrease progressively the electronic correlations. This may, in turn, change the balance between various fluctuations and pairing instabilities. Nevertheless, our findings demonstrate that the pairing interactions are sensitive to chemical doping. The observation of tunable pairing symmetry may also reconcile the previously contradictory data regarding the pairing symmetry in iron based materials [12,18,29,30].

We thank D. L. Feng and X. H. Niu for helpful discussions. This work is supported by the National Natural Science Foundation of China (No. 11374059) and the Ministry of Science and Technology of China (973 project: 2015CB921302). Use of the high flux isotope reactor at the Oak Ridge National Laboratory was supported by the U.S. Department of Energy, Office of Basic Energy Sciences, Scientific User Facilities Division. The research at UC Berkeley is supported by the Director, Office of Science, Office of Basic Energy Sciences, U.S. Department of Energy, under Contracts No. DE-AC02-05CH11231, No. DE-AC03-76SF008, and No. DE-AC02-05CH11231.

*jitae.park@frm2.tum.de

† zhaoj@fudan.edu.cn

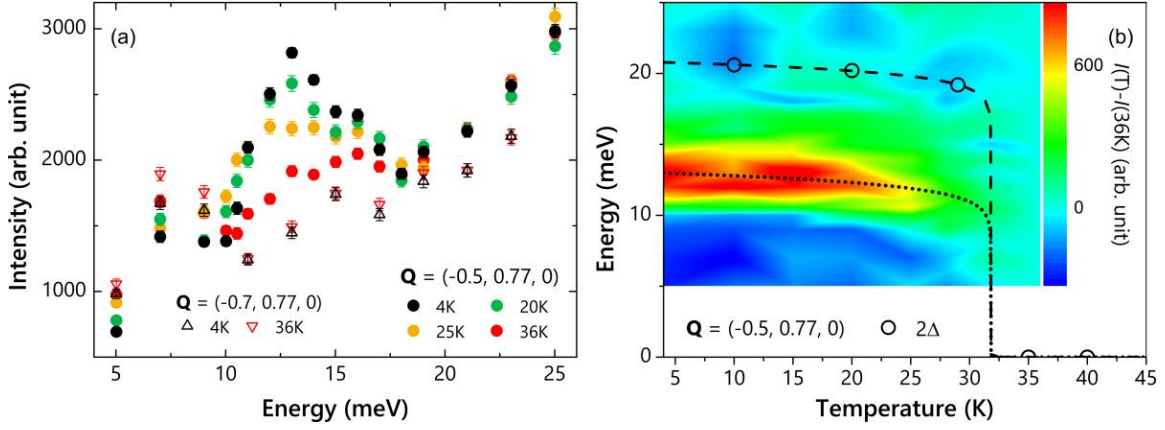

FIG. 1. The Energy dependence of the magnetic excitations and their temperature evolution in $K_xFe_{2-y}Se_2$. (a) Energy scans at **Q** = (-0.5, 0.77, 0) and **Q** = (-0.7, 0.77, 0) for temperatures below and above $T_c$. A superconductivity-induced resonant mode is clearly observed at 13 meV at **Q** = (-0.5, 0.77, 0), while the scattering at **Q** = (-0.7, 0.77, 0) displays little temperature dependence across $T_c$. (b) Background subtracted contour map of change in the magnetic scattering as a function of energy and temperature. The background was measured at $T$ = 36 K. The open circles indicate 2Δ adapted from the ARPES measurements of ref. [43], which is consistent with the data measured on a single crystal from the same batch used for our neutron scattering measurements (not shown). The dotted line and dashed line are guides to the eye. The error bars indicate one standard deviation throughout the Letter.

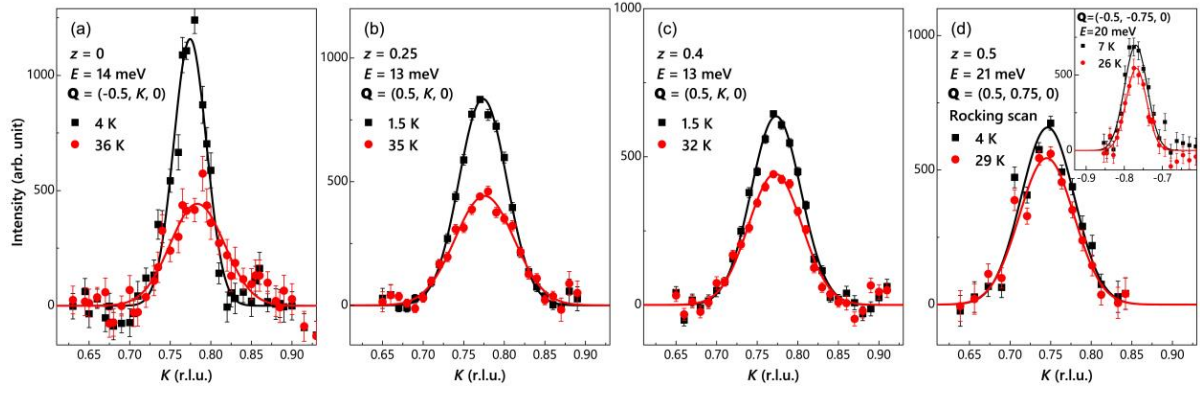

FIG. 2. Momentum dependence of the magnetic excitations in $K_xFe_{2-y}(Se_{1-z}S_z)_2$ ($z = 0, 0.25, 0.4, 0.5$) below and above $T_c$. (a)-(d) Constant energy scans in the superconducting (black) and the normal state (red) for the $z = 0, 0.25, 0.4, 0.5$ samples, respectively. The energies were chosen to correspond to the peak in the extra scattering associated with the superconductivity. The data fit well to Gaussian curves as indicated by solid lines.

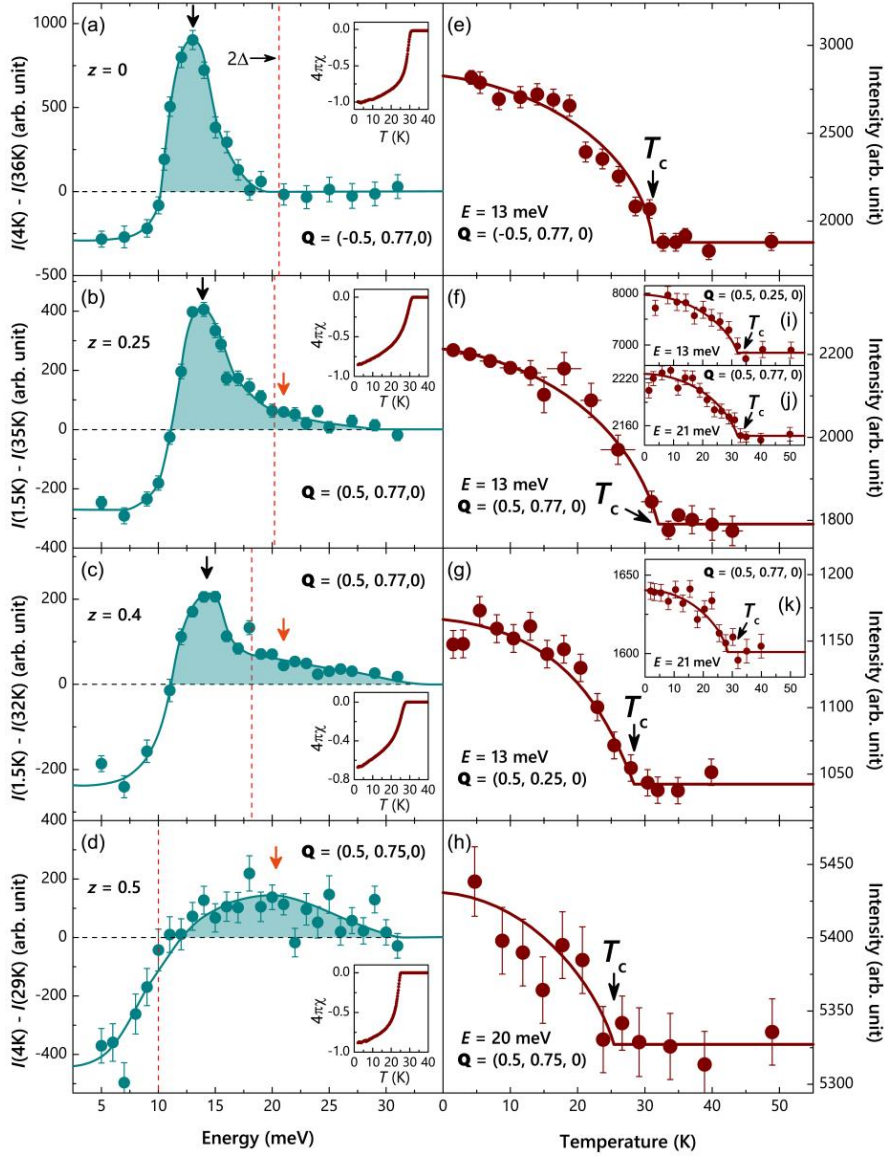

FIG. 3. Temperature dependence of the spin excitations and their doping evolution in $K_xFe_{2-y}(Se_{1-z}S_z)_2$ ($z$ = 0, 0.25, 0.4, 0.5) below and above $T_c$. (a)-(d) Energy dependence of the intensity difference between the superconducting and the normal states in the vicinity of **Q** = (0.5, 0.75, 0) or equivalent wave vectors. The dashed lines represent 2Δ determined by ARPES measurements [33]. The shaded area denotes the spectral weight enhancement below $T_c$. The resonance energies and hump/shoulder energies are marked by black and orange arrows, respectively. The data are normalized to the amplitude of the normal state **Q**-scans at similar energies between different doping levels. The insets show the temperature dependence of DC magnetic susceptibility. The magnetic field ($H$ = 10 Oe) is applied parallel to the $ab$ plane. 1 Oe = (1000/4π) A/m. (e)-(k) Temperature dependence of the spin excitations measured at various energies. (e) $z$ = 0, $E$ = 13 meV, **Q** = (-0.5, 0.77, 0); (f) $z$ = 0.25, $E$ = 13 meV, **Q** = (0.5, 0.77, 0); (g) $z$ = 0.4, $E$ = 13 meV, **Q** = (0.5, 0.25, 0); (h) $z$ = 0.5, $E$ = 20 meV, **Q** = (0.5, 0.75, 0); (i) $z$ = 0.25, $E$ = 13 meV, **Q** = (0.5, 0.25, 0); (j) $z$ = 0.25, $E$ = 21 meV, **Q** = (0.5, 0.77, 0); (k) $z$ = 0.4, $E$ = 21 meV, **Q** = (0.5, 0.77, 0). The horizontal error bars indicate the temperature range within which data were combined. The solid lines are guides to the eye.

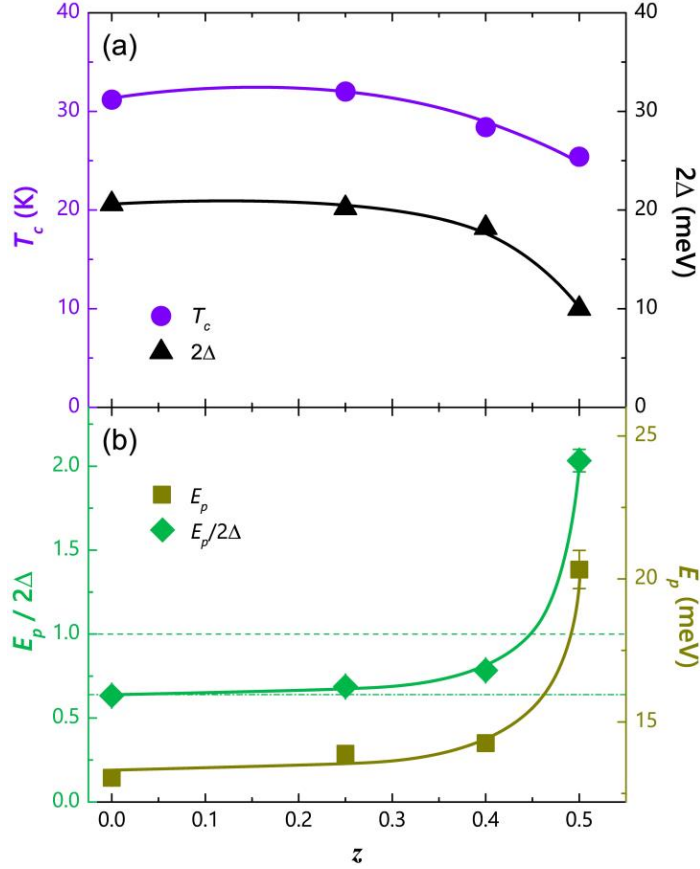

FIG. 4. Doping dependence of $T_c$, superconducting gap ($2\Delta$), and superconductivity induced magnetic peak energy ($E_p$). (a) Doping dependence of the $T_c$ and the superconducting gap ($2\Delta$) adapted from ARPRES measurements in ref. [33]. (b) Doping dependence of the superconductivity-induced magnetic peak energy ($E_p$) and the ratio of $E_p/2\Delta$. For $z = 0$, 0.25, and 0.4 samples, $E_p$ denotes the resonance energy. For $z = 0.5$ sample, $E_p$ denotes the energy of the hump structure above $2\Delta$. The dot-dashed line indicates the empirical universal ratio of $E_p/2\Delta = 0.64$ for magnetic superconductors [35]. The dashed line indicates $E_p/2\Delta = 1$.